\documentstyle[12pt]{article}

\begin{document}

\begin{center}
\baselineskip 40pt

\vskip 2cm

{\Large {\bf Dynamics of a Two-Level System Coupled to Ohmic Bath:
A Perturbation Approach
}}

\vskip 1cm

{\large {\rm H.Zheng}}

Department of Physics, Shanghai Jiao Tong University \\
Shanghai 200030, People's Republic of China \\

{\bf Abstract} 
\end{center}

\baselineskip 20pt 

The physics of a two-level system coupled to Ohmic bath is studied 
by means of the perturbation approach based on a unitary transformation.
Our main results are: The coherence-incoherence transition is at
$\alpha_c={1\over 2}[1+\Delta_r/\omega_c]$; for $\alpha<\alpha_c$
the dynamical quantity $P(t)=\cos(\omega_0 t)\exp(-\gamma t)$; the
sussceptibility $\chi^{\prime\prime}(\omega)/\omega$ is of a double peak
structure for $\alpha<\alpha_c$ and the Shiba's relation is exactly
satisfied; at the transition point $\alpha=\alpha_c$ the real time
correlation function $C(t)\approx -1/\gamma^2_c t^2$ in the long time limit. 

\vskip 1cm 


\pagebreak 

\baselineskip 20pt 

Studies of the two-level system coupled to Ohmic bath (spin-boson model, SBM) 
have a long history\cite{rmp,book}. Although there were many studies using various 
kinds of methods, in recent years the physics of SBM has still attracted 
considerable attention because it provides a universal model for many physical 
systems\cite{rmp,book}. The Hamiltonian of SBM reads
\begin{equation}
H=-{\frac{1}{2}}\Delta \sigma _{x}
+\sum_{k}\omega _{k}b_{k}^{\dag }b_{k}+{\frac{1}{2}}
\sum_{k}g_{k}(b_{k}^{\dag }+b_{k})\sigma _{z}.
\end{equation}
The notations are the same as usual\cite{rmp,book}. In this work we consider the
zero bias case with temperature $T=0$. The Ohmic bath is characterized by 
its spectral density: 
$\sum_{k}g^2_{k}\delta(\omega-\omega_k)=2\alpha\omega\theta(\omega_c-\omega)$,
where $\alpha$ is the dimensionless coupling constant and $\theta(x)$ is the
usual step function.

The Hamiltonian (1) seems to be quite simple. However, it cannot be solved 
exactly and various approximate analytical and numerical methods have been
used[1-13]. Recent studies of SBM focused on its dynamical
properties[6-12]. Among these studies, the variational approach proposed 
by Silbey and Harris\cite{sh} is of interest. They proposed to make a unitary 
transformation: $H^{\prime}=\exp(S)H\exp(-S)$, with
\begin{eqnarray}
&&S=\sum_{k}\frac{g_{k}}{2\omega _{k}}\xi _{k}(b_{k}^{\dag }-b_{k})\sigma_z.
\end{eqnarray}
Then, the variational ground state energy is: 
$E_g=\langle s_1|\langle \{0_{k}\}|H'|s_1\rangle |\{0_{k}\}\rangle$,
where $|s_{1}\rangle ={1\over\sqrt{2}}\left( 
\begin{array}{c}
1 \\ 
1
\end{array}
\right) $ and $|s_{2}\rangle ={1\over\sqrt{2}}\left( 
\begin{array}{c}
1 \\ 
-1
\end{array}
\right) $ are eigenstates of $\sigma_x$ and $|\{0_{k}\}\rangle$ is
the vacuum state of bosons in which $n_{k}=0$ for every $k$. 
By minimizing $E_g$ $\xi _{k}$ is determined as
\begin{eqnarray}
\xi _{k} &=&\omega _{k}/(\omega _{k}+\eta\Delta).
\end{eqnarray}
The authors calculated the renormalized tunnelling $\Delta_r$\cite{sh},
\begin{eqnarray}
&&\eta=\frac{\Delta_r}{\Delta}=\left(e\frac{\Delta}{\omega_c}\right)^{\frac{\alpha}{1-\alpha}}
  \exp\left[-\frac{\alpha}{1-\alpha}\left(\frac{\eta\Delta}{\omega_c+\eta\Delta}
  +\ln\frac{\omega_c+\eta\Delta}{\omega_c}\right)\right].
\end{eqnarray}
Although the variational method can predict correctly the
delocalized-localized transition point $\alpha_l=1$ at the scaling limit
$\Delta/\omega_c\ll 1$, it is not suitable for calculating the dynamical
properties of SBM. In this work we present a new analytical approach,
based on the unitary transformation ((2) and (3)) and the perturbation
theory\cite{zhe}, for calculating the dynamical preperties of SBM which 
works well for the coupling constant $0<\alpha<1$ and the bare tunnelling
$0<\Delta<\omega_c$. The approach is quite simple and physically clear, and
may be easily extended to more complicated coupling systems. Throughout this 
work we set $\hbar =1$ and $k_{B}=1$.

The unitary transformation (2) can be done to the end and the result is 
$H'=H'_{0}+H'_{1}+H'_{2}$, where
\begin{eqnarray}
&&H_{0}^{\prime }=-{\frac{1}{2}}\eta \Delta \sigma _{x}
+\sum_{k}\omega _{k}b_{k}^{\dag }b_{k} 
-\sum_{k}\frac{g_{k}^{2}}{4\omega _{k}}\xi _{k}(2-\xi _{k}), \\
&&\eta =\exp [-\sum_{k}\frac{g_{k}^{2}}{2\omega _{k}^{2}}\xi _{k}^{2}],\\
&&H_{1}^{\prime }={\frac{1}{2}}\sum_{k}g_{k}(1-\xi _{k})(b_{k}^{\dag
}+b_{k})\sigma _{z}-{\frac{1}{2}}\eta \Delta i\sigma_{y}
\sum_{k}\frac{g_{k}}{\omega _{k}}\xi _{k}(b_{k}^{\dag }-b_{k}),  \\
&&H_{2}^{\prime }=-{\frac{1}{2}}\Delta\sigma_{x}\left( \cosh
\{\sum_{k}\frac{g_{k}}{\omega _{k}}\xi _{k}(b_{k}^{\dag }-b_{k})\}-\eta
\right)   \nonumber \\
&&-{\frac{1}{2}}\Delta i\sigma_{y}\left( \sinh \{\sum_{k}\frac{g_{k}}{%
\omega _{k}}\xi _{k}(b_{k}^{\dag }-b_{k})\}-\eta \sum_{k}\frac{g_{k}}{%
\omega _{k}}\xi _{k}(b_{k}^{\dag }-b_{k})\right) .
\end{eqnarray}
Obviously, $H_{0}^{\prime }$ can be solved exactly because in
which the spin and bosons are decoupled. 
The eigenstate of $H_{0}^{\prime}$ is a direct
product, $|s\rangle |\{n_{k}\}\rangle$, where $|s\rangle $ is $|s_{1}\rangle$
or $|s_{2}\rangle$ and $|\{n_{k}\}\rangle$ means that there are $n_{k}$
phonons for mode $k$. The ground state of $H_{0}^{\prime}$ is
\begin{equation}
|g_{0}\rangle =|s_{1}\rangle |\{0_{k}\}\rangle . 
\end{equation}
$H_{1}^{\prime }$ and $H_{2}^{\prime}$ are treated as perturbation and
they should be as small as possible. Eq.(3),
$\xi_{k}=\omega_{k}/(\omega _{k}+\eta\Delta)$, leads to
\begin{eqnarray}
&&H_{1}^{\prime}=
\frac{1}{2}\eta \Delta\sum_{k}\frac{g_{k}}{\omega _{k}}\xi _{k}
\left[b_{k}^{\dag }(\sigma_z-i\sigma_y)+b_{k}(\sigma_z+i\sigma_y)
\right] 
\end{eqnarray}
and $H_{1}^{\prime}|g_{0}\rangle =0$.
Note that the form of $\xi_k$, Eq.(3), in this work is determined by
$H_{1}^{\prime}|g_{0}\rangle =0$, instead of minimizing the ground state
energy $E_g$. Thus, we get rid of the variational condition.

The lowest excited states are $|s_2\rangle |\{0_{k}\}\rangle$ and 
$|s_1\rangle |1_k \rangle$, where $|1_k \rangle$ is the number state
with $n_k=1$ but $n_{k'}=0$ for all $k'\neq k$. It's easily to check that
$\langle g_0|H'_2|g_{0}\rangle =0$ (because of the form of $\eta$ in Eq.(6)),
$\langle\{0_{k}\}|\langle s_2|H'_2|g_{0}\rangle =0$, 
$\langle 1_{k}|\langle s_1|H'_2|g_{0}\rangle =0$, and
$\langle\{0_{k}\}|\langle s_2|H'_2|s_1\rangle |1_k \rangle=0$.
Moreover, since $H_{1}^{\prime}|g_{0}\rangle =0$, we have
$\langle\{0_{k}\}|\langle s_2|H'_1|g_{0}\rangle =0$ and
$\langle 1_{k}|\langle s_1|H'_1|g_{0}\rangle =0$. Thus, we can diagonalize
the lowest excited states of $H'$ as 
\begin{eqnarray}
&&H'=-\frac{1}{2}\eta \Delta |g_{0}\rangle\langle g_0|
     +\sum_E E |E\rangle\langle E|+\mbox{terms with higher excited states}.
\end{eqnarray}
The diagonalization is through the following transformation\cite{gui}:
\begin{eqnarray}
&&|s_2\rangle |\{0_{k}\}\rangle=\sum_E x(E)|E\rangle,\\
&&|s_1\rangle |1_k \rangle=\sum_E y_k(E)|E\rangle,\\
&&|E\rangle=\sum_E x(E)|s_2\rangle |\{0_{k}\}\rangle
+\sum_E y_k(E)|s_1\rangle |1_k \rangle,
\end{eqnarray} 
where
\begin{eqnarray}
&&x(E)=\left[1+\sum_k\frac{V^2_k}{(E+\eta\Delta/2-\omega_k)^2}\right]^{-1/2},\\
&&y_k(E)=\frac{V_k}{E+\eta\Delta/2-\omega_k}x(E),
\end{eqnarray}
with $V_k=\eta\Delta g_k\xi_k/\omega_k$. $E$'s are the diagonalized 
excitation energy and they are solutions of the equation 
\begin{eqnarray}
&&E-\frac{\eta\Delta}{2}-\sum_k\frac{V^2_k}{E+\eta\Delta/2-\omega_k}=0.
\end{eqnarray}

The dynamical quantity $P(t)=\langle b,+1|\langle +1|
e^{iHt}\sigma _{z} e^{-iHt}|+1\rangle |b,+1\rangle$ is defined in
Ref.\cite{rmp}, where $|+1\rangle$ is the eigenstate of
$\sigma_z=+1$ and $|b,+1\rangle$ is the state of bosons adusted to
the state of $\sigma_z=+1$. Because of the unitary transformation
($e^S\sigma_z e^{-S}=\sigma_z$)
\begin{eqnarray}
&&P(t)=\langle \{0_{k}\}|\langle +1|
e^{iH't}\sigma _{z} e^{-iH't}|+1\rangle |\{0_{k}\}\rangle,
\end{eqnarray}
since $e^S|+1\rangle |b,+1\rangle=|+1\rangle |\{0_{k}\}\rangle$.
Using Eqs.(11)-(17) the result is
\begin{eqnarray}
&&P(t)={1\over 2}\sum_E x^2(E)\exp[-i(E+\eta\Delta/2)t]
+{1\over 2}\sum_E x^2(E)\exp[i(E+\eta\Delta/2)t] \nonumber\\
&&=\frac{1}{4\pi i}\oint_C dE'e^{-iE't}
\left(E'-\eta\Delta-\sum_k\frac{V^2_k}{E'+i0^+-\omega_k}\right)^{-1}\nonumber\\
&&+\frac{1}{4\pi i}\oint_C' dE'e^{iE't}
\left(E'-\eta\Delta-\sum_k\frac{V^2_k}{E'-i0^+-\omega_k}\right)^{-1},
\end{eqnarray}
where a change of the variable $E'=E+\eta\Delta/2$ is made.
The real and imaginary parts of $\sum_kV^2_k/(E'\pm i0^+-\omega_k)$
are denoted as $R(E')$ and $\mp\gamma(E')$,
\begin{eqnarray}
&&R(\omega)=-2\alpha\frac{(\eta\Delta)^2}{\omega+\eta\Delta}
\left\{\frac{\omega_c}{\omega_c+\eta \Delta}
-\frac{\omega}{\omega+\eta \Delta}
\ln\left[\frac{|\omega|(\omega_c+\eta \Delta)}
{\eta\Delta(\omega_c-\omega)}\right]\right\},\\
&&\gamma(\omega)=2\alpha\pi\omega(\eta\Delta)^2/(\omega+\eta\Delta)^2.
\end{eqnarray}
The integral in (19) can proceed by calculating the residue of integrand
and the result is $P(t)=\cos(\omega_0 t)\exp(-\gamma t)$, where
$\omega_0$ is the solution of equation
\begin{eqnarray}
\omega-\eta\Delta-R(\omega)=0
\end{eqnarray}
and $\gamma=\gamma(\eta\Delta)=\alpha\pi\eta\Delta/2$
(the second order approximation).
This $P(t)$ is of the form of damped oscillation and
one can check that the solution $\omega_0$ is real when
$1>2\alpha\omega_c/(\omega_c+\eta \Delta)$. When
$1<2\alpha\omega_c/(\omega_c+\eta \Delta)$, the solution
$\omega_0$ is imaginary and we have an incoherent $P(t)$.
$\alpha_c={1\over 2}[1+\eta\Delta/\omega_c]$ determines
the critical point where there is a coherent-incoherent transition.
Note that when $\Delta/\omega_c\ll 1$, we have $\alpha_c=1/2$,
$\omega_0=0$ and $P(t)=\exp(-\gamma_c t)$ ($\gamma_c=\pi e\Delta^2/4\omega_c$
since $\eta=e\Delta/\omega_c$ from Eq.(4)),
which is the same as was predicted by previous authors
($\gamma_c=\pi\Delta^2/2\omega_c$ in Ref.\cite{wei,gui,cr}).
Fig.1 shows the calculated $\omega_0/\Delta_r$
as functions of $\alpha$ ($\alpha\le\alpha_c$) for $\Delta=0.01$, $0.1$,
and $0.5$. $\Delta_r=\eta\Delta$ is the renormalized tunnelling.
The dotted line is $\gamma/\Delta_r=\alpha\pi/2$. 

Since $e^S\sigma_z e^{-S}=\sigma_{z}$, the retarded Green's function is
\begin{eqnarray}
G(t)&=&-i\theta (t)\left\langle [\exp (iH^{\prime }t)\sigma _{z}\exp
(-iH^{\prime }t),\sigma _{z}]\right\rangle ^{\prime },
\end{eqnarray}%
where $\langle ...\rangle ^{\prime }$ means the average with
thermodynamic propability $\exp (-\beta H^{\prime })$.
The Fourier transformation of $G(t)$ is denoted as
$G(\omega)$, which satisfies an infinite chain of equation of
motion\cite{mah}. We have made the cutoff approximation for the
equation chain at the second order of $g_{k}$ and the solution at $T=0$ is
\begin{eqnarray}
&&G(\omega)=\frac{1}{\omega -\eta\Delta -\sum_{k}V^2_k/(\omega -\omega _{k})} 
-\frac{1}{\omega +\eta\Delta -\sum_{k}V^2_k/(\omega +\omega _{k})}.
\end{eqnarray}
The susceptibility $\chi(\omega)=-G(\omega)$, and its imaginary part is
\begin{eqnarray}
&&\chi^{\prime\prime}(\omega)=
\frac{\gamma(\omega)\theta(\omega)}
{[\omega -\eta \Delta -R(\omega)]^2+\gamma^2(\omega)} 
-\frac{\gamma(-\omega)\theta(-\omega)}
{[\omega +\eta \Delta +R(-\omega)]^2+\gamma^2(-\omega)}.
\end{eqnarray}

The $\omega\to 0$ limit of $S(\omega)=\chi^{\prime\prime}(\omega)/\omega$ is 
\begin{eqnarray}
&&\lim_{\omega\to 0}\frac{\chi^{\prime\prime}(\omega)}{\omega}
=\frac{2\alpha\pi}
{(\eta \Delta)^2\{1-2\alpha[1-\eta \Delta/(\omega_c+\eta \Delta)]\}^2}.
\end{eqnarray}
Besides, the real part of the susceptibility is
\begin{eqnarray}
&&\chi^{\prime}(\omega=0)=\frac{2}
{\eta \Delta \{1-2\alpha[1-\eta \Delta/(\omega_c+\eta \Delta)]\}}.
\end{eqnarray}
Thus, the Shiba's relation\cite{wei,vol,cos,ks}
\begin{equation}
\lim_{\omega\to 0}\frac{\chi^{\prime\prime}(\omega)}{\omega}=
{\pi\over 2}\alpha\chi^{\prime}(\omega=0)^2
\end{equation}
is exactly satisfied. $S(\omega)$ has a double peak structure for
$\alpha<\alpha_c$. For $\alpha\ge \alpha_c$ there is only one peak at
$\omega=0$. Fig.2 shows the $S(\omega)$ versus $\omega$ relations for
fixed $\alpha=0.3$ and $\Delta=0.01$, $0.1$, and $0.5$.

The symmetrized correlation function
\begin{eqnarray}
&&C(t)=\frac{1}{2}\mbox{Tr}\left\{\exp(-\beta H)[\sigma _{z}(t)\sigma_z
+\sigma_z\sigma_z(t)]\right\} /\mbox{Tr}[\exp (-\beta H)]\nonumber\\
&&=-\frac{1}{2\pi}\int^{\infty}_{-\infty}d\omega\coth(\frac{\beta\omega}{2})
\mbox{Im}G(\omega)\exp(-i\omega t)\nonumber\\
&&=\frac{1}{\pi}\int^{\infty}_0 d\omega
\frac{\gamma(\omega)}
{[\omega -\eta \Delta -R(\omega_0)]^2+\gamma^2}\cos(\omega t),
\end{eqnarray}
where $\omega_0$ is the solution of Eq.(22) and $\gamma(\omega)$ in the
denominator is approximated by the second order approximation $\gamma$.
At the scaling limit $\Delta/\omega_c\ll 1$ and the coherence-incoherence
transition point $\alpha=1/2$,
\begin{eqnarray}
&&C(t)=\int^{\infty}_0 d\omega
\frac{\omega\cos(\omega t)}{\omega^2+\gamma^2_c}
\frac{(\eta\Delta)^2}{(\omega+\eta\Delta)^2}.
\end{eqnarray}
$C(t)$ decays algebraically in the long-time limit: $C(t)\approx -1/\gamma^2_c t^2$, 
which is the same as what was predicted by previous authors.

In summary: The physics of SBM is studied by means of the perturbation
approach based on a unitary transformation. Analytical results of
the dynamical quantity $P(t)$, the sussceptibility
$\chi^{\prime\prime}(\omega)$ and the real time correlation function $C(t)$
are obtained for both the scaling limit $\Delta_r/\omega_c\ll 1$ and the general finite
$\Delta_r/\omega_c$ case. Our approach is quite simple, but it can reproduce
nearly all results which agree with those of previous authors using various complicated 
methods. Besides, our approach can be easily extended to other more complicated 
coupling systems.

\vskip 0.5cm 

{\noindent {\large {\bf Acknowledgement}}}

This work was supported by the China National Natural Science Foundation 
(Grants No.10074044 and No.90103022). 

\rm 

\newpage

\begin{center}
{\Large \bf Figure Captions }
\end{center}

\vskip 0.5cm

\baselineskip 20pt

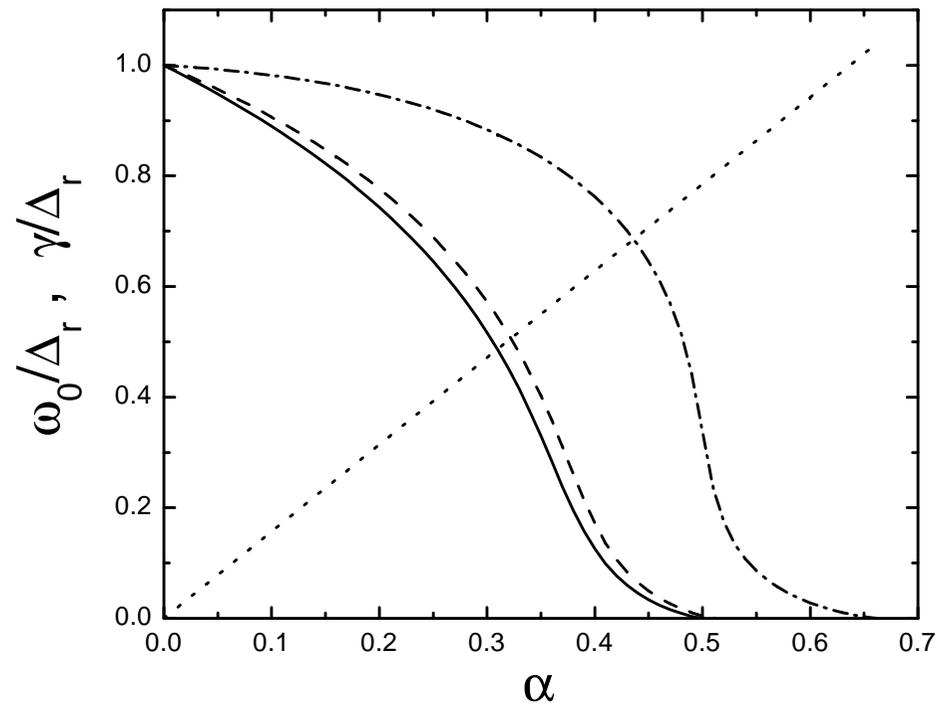
{\bf Fig.1}~~~$\omega_0/\Delta_r$ versus $\alpha$ relations for 
     $\Delta/\omega_c=0.01$ (solid line, $\alpha_c=0.50014$), 
     $0.1$ (dashed, $\alpha_c=0.51212$), and
     $0.5$ (dashed-dotted, $\alpha_c=0.66244$). The dotted line
     is $\gamma/\Delta_r=\alpha\pi/2$.

\vskip 0.5cm 

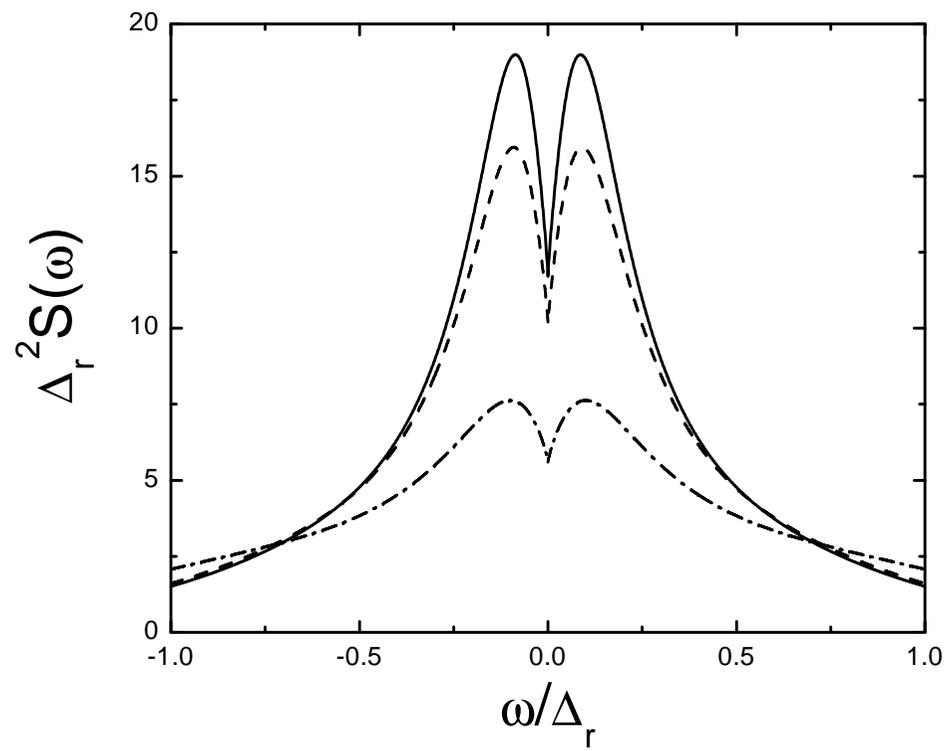
{\bf Fig.2}~~~$S(\omega)$ as functions of $\omega$ for fixed
     $\alpha=0.3$ and $\Delta=0.01$ (solid line), $0.1$ (dashed), 
     and $0.5$ (dashed-dotted).

\end{document}